\documentclass[pdflatex]{sn-jnl}

\usepackage{graphicx}% Include figure files
\usepackage{siunitx}
\usepackage{amsmath}
\usepackage{ulem} % cross out text
\usepackage[utf8]{inputenc}
\usepackage{xspace}
\usepackage{afterpage}
\usepackage{lipsum}
\usepackage{url}
\usepackage{cite}

\usepackage{nicefrac}
\usepackage{changes} % lets one use \added{} \replaced{}{} and \deleted{}
\graphicspath{{figure/}}  
\newcommand{\abs}[1]{\left|#1\right|}
\newcommand{\rbrac}[1]{\left(#1\right)}
\newcommand{\kB}[0]{k_\text{B}}
\newcommand{\psiL}{\psi_\text{L}}
\newcommand{\psiR}{\psi_\text{R}}
\newcommand{\psiA}{\psi_\text{A}}
\newcommand{\psiS}{\psi_\text{S}}

\renewcommand{\rm}[1]{{}}

\title{3D micro-printing: An enabling technique for arbitrary potential landscapes for photonic quantum-gases}

\author*[1]{\href{https://orcid.org/0000-0003-4630-4117}{\fnm{Julian}\sur{Schulz}}}\email{schulzj@rptu.de}
\author[2]{\href{https://orcid.org/0000-0002-3644-1233}{\fnm{Kirankumar}\sur{Karkihalli Umesh}}}
\author[1]{\href{https://orcid.org/0009-0000-8092-5676}{\fnm{Sven}\sur{Enns}}}
\author[2]{\href{https://orcid.org/0000-0001-7818-2981}{\fnm{Frank}\sur{Vewinger}}}
\author[1,3]{\href{https://orcid.org/0000-0003-2389-5532}{\fnm{Georg}\sur{von Freymann}}}

\affil[1]{\orgdiv{Physics Department and Research Center OPTIMAS}, \orgname{RPTU University Kaiserslautern Landau}, \orgaddress{\city{Kaiserslautern}, \postcode{67663}, \country{Germany}}}

\affil[2]{\orgdiv{Institut für Angewandte Physik}, \orgname{Universität Bonn}, \orgaddress{\street{Wegelerstrasse 8}, \city{Bonn}, \postcode{53115}, \country{Germany}}}

\affil[3]{\orgname{Fraunhofer Institute for Industrial Mathematics ITWM}, \orgaddress{\city{Kaiserslautern}, \postcode{67663}, \country{Germany}}}

\date{\today}% It is always \today, today,
\begin{document}
% currently 195 words
% Nature Nano wants 150

\abstract{
%The behaviour of quantum gases in different potential landscapes has been of wide interest for investigating solid-state physics, transport phenomena, quantum information, or precision measurements. 
%While these systems are often studied by cooling gases and exposing them to optical lattices, they can also be replicated on a photonic platform under ambient conditions by using, for example, an optical dye-filled micro-cavity to condense a gas of photons into the ground state. 
%\fvc{There is no mentioning of thermodynamics. We should stress this, as people usually think about waveguides when talking about photonic systems.}
%\js{In such dye-filled micro-cavities, thermodynamic properties and quantum phase transitions, like Bose-Einstein-Condensation can be studied.}
%Here we employ direct laser writing (DLW) as a structuring technique to fabricate polymer structures on one of the cavity mirrors that act as a potential landscape for the photon gas.
%DLW allows us to create arbitrary potential landscapes with a feature size that is multiple orders of magnitude smaller than previously reported structuring techniques, thus allowing us to archive steeper potentials and correspondingly multiple orders of magnitude higher coupling rates, enabling the study of large lattice structures.
%We demonstrate the versatility of this structuring technique for photon gases in a cavity with a box potential, an anisotropic harmonic potential, a double-well potential, and a topological non-trivial lattice.
Photonic quantum gases explore the physics of open driven-dissipative quantum systems under ambient conditions and thus open access to thermodynamics and transport phenomena in quantum gases in the weakly interacting regime. Here we introduce the technology of 3D micro-printing to create potential landscapes for photonic quantum gases in dye-filled micro cavities, which surpass the current state of the art in terms of potential size and definition, potential depth, coupling strength, and number of coupled potentials by at least an order of magnitude. We realize as demonstration of the capabilities box potentials with rectangular side walls, anisotropic harmonic potentials, double-well potentials with dimensions on the scale of the wavelength of light as well as potential lattices with topological non-trivial properties. This approach paves the way for experimentally studying the physics of open quantum systems on lattices and might find applications in solving complex ground-state problems like the XY-model.}


\maketitle

% Nature Nano wants 3000 words Main text excluding abstract, Methods, references and figure legends.
% currently ~2950 words (figure legends==figure captions?)

%{\color{red} Please take also a look at the cover letter (compile Coverletter.tex).}

\section{Introduction}

% Intruduction
The ideal Bose gas describes an ensemble of non-interacting bosons in thermal equilibrium. 
At low temperatures or high densities, the Bose gas undergoes a phase transition to a Bose-Einstein condensate (BEC), which is characterized by a macroscopic coherent occupation of the ground state. 
This phase transition has been observed not only in cold atomic gases but also in quasiparticles such as exciton-polaritons \cite{deng_review,carusotto_review} or magnons \cite{Rezende2020} and has also been experimentally realized with photons confined in a dye-filled cavity \cite{klaers.2010,Marelic.2015}.
The energy of a cavity mode $\psi$ can be approximated in the paraxial limit ($k_r\ll k_z$) with the equation\cite{vretenar_modified_2021}:
\begin{equation}
    E=\rbrac{m_\text{ph}c_n^2
    +\frac{\hbar^2k_r^2}{2m_\text{ph}}
    -V(x,y)},
    \label{eq:eigenenergy}
\end{equation}
where $\hbar$ is the reduced Planck constant, $c_n=c/n$ is the speed of light in the dye solution that has a refractive index of $n$ and $m_\text{ph}$ is the effective photon mass.
The effective photon mass is given by the cut-off frequency of the cavity $\omega_\text{cut}$ that depends on the mechanical length of the cavity $D_0$ and the longitudinal mode number $q$.
The first term in Eq.~\eqref{eq:eigenenergy} is the rest mass energy or the ground state energy in the cavity $E_\text{ground}=\hbar\omega_\text{cut}=\hbar c_n\frac{q\pi}{D_0}=m_\text{ph}c_n^2$.
The second term in Eq.~\eqref{eq:eigenenergy} is the kinetic energy, where $k_r$ is the transverse component of the wave vector.
The longitudinal wavevector $k_z$ perpendicular to the cavity mirror surface can be assumed as fixed, as the short cavity imposes $k_z \gg k_r$.
The last term in Eq.~\eqref{eq:eigenenergy} is the potential experienced by the photon gas. 

% thermal structuring technique
In recent years, various techniques have been developed to produce variable potentials for such photon gases.
The mirror surface can be deformed by the thermal expansion of a layer directly under the Bragg layers \cite{Kurtscheid.2019,Kurtscheid.2020,Busley.2022, vretenar_structuring_2023}, or the refractive index can be changed locally by heating a temperature-sensitive polymer in the cavity \cite{Dung.2017}, or a combination of both \cite{vretenar_modified_2021, vretenar_josephson_2021}.
Because these techniques are based on thermal effects, they are limited in their spatial resolution to a few micrometers, which is why they currently cannot be used to, for example, generate deep potentials or potentials with edges sharper than the wavelength of the photon gas.
% milling as a structuring technique
Another approach is to first mill the potential with a CO$_2$ laser or an ion beam into the substrate and then add the Bragg layer coatings in a second step \cite{Trichet:15, Walker:21, Hunger_2010}. While the ion beam milling allows fabricating steep structures, the subsequent coating limits the spatial resolution, since the Bragg layers can only be bent by a small amount before the layers crack and greatly reduce the finesse of the mirror cavity.

% In this paper
In this paper, we employ 3D micro-printing (also known as direct laser writing, DLW) to create highly-defined polymer micro-/nano-structures on the cavity mirror surface. 
% DLW
DLW is a lithography method based on two-photon absorption, which is widely employed in the fields of, e.g., 3D photonic crystals, 3D metamaterials, microoptics, bio-templating (see review and recent primer articles \cite{Hohmann.2015,Skliutas.2015} for more details). A UV-sensitive negative tone photoresist is applied on a substrate and an infrared femtosecond pulsed laser is focused into it via an objective. Only in the focal volume is the field strength high enough to excite a photoinitiator molecule via two-photon absorption and initialize a polymerization reaction in the photoresist. 
By moving the focal volume relative to the substrate, arbitrary chosen volumes of the photoresist can be exposed. Thus, the photoresist hardens to a solid polymer structure.
The unexposed photoresist is subsequently washed away, leaving a 3D micro-printed polymer structure on the substrate.

\section{Results}

% basics of photon BEC
To observe the phase transition of a photon gas into a BEC state, thermalization is realized in a micro-cavity filled with a dye solution (Rhodamine 6G in ethylene glycol). 
The dye molecules are initially excited with an external pump laser (wavelength \SI{532}{\nano\meter}). Most of the spontaneously emitted photons populate the cavity modes. The highly reflective cavity mirrors hold the photons long enough to thermalize to room temperature via absorption and re-emission cycles via the dye molecules. 
Since the cavity mirrors are separated by only a few micrometers, the emission spectrum of the dye overlaps only with one longitudinal mode of the cavity, thus defining a ground state with non-zero energy and, therefore, creating an effective photon mass.
Because only one longitudinal mode can be occupied, this photon gas is effectively two-dimensional \cite{klaers.2010}. 
By locally changing the optical path length between the mirrors, the photon gas can experience a predetermined potential.

% sample preparation
Here, we use specifically coated dielectric mirrors for the photon BEC micro-cavity as a substrate on which the polymeric structures forming the potentials are printed using DLW. 
The mirror coating allows the formation of a high-finesse cavity with a finesse of about $10^5$ for the empty cavity at the wavelength range covered by the photon gas around $\lambda\in[\SI{570}{\nano\meter},\SI{600}{\nano\meter}]$, while transmitting more than $80\%$ of the light of the pump laser with $\lambda=\SI{532}{\nano\meter}$ at an angle of incidence of~\SI{0}{\degree}. 
After the DLW-process the structured cavity mirrors are placed in a common photon BEC setup where upon pumping the photons populate the modes of the potential imposed by DLW. By imaging the radiation emitted through one of the cavity mirrors, we can analyze the distribution of photons within the potential \cite{karkihalli_umesh_dimensional_2024}.

% theoretical energy spectrum

The potential for the photon gas depends on the height profile of the polymer structure $h_\text{s}=(x,y)$. Depending on the refractive index difference $\Delta n$ between the polymer and the dye solution, the polymer structures can act as attractive or repulsive potentials for the photon gas.
The resulting potential for the photon gas is given by
\begin{equation}
    V(x,y)=m_\text{ph}c_n^2\frac{h_\text{s}\rbrac{x,y}\Delta n}{D_0 n}.
\end{equation}
Due to the intrinsically high lateral resolution of DLW, potentials with sharp edges well below the wavelength of the photons in the quantum gas can be created.
These polymer structures do not deform or damage the underlying Bragg layers, preserving the high cavity finesse. Furthermore, they can be removed to reuse and restructure these mirrors.
In our case, the polymer has a refractive index higher than that of the dye solution, so $\Delta n$ is positive and $V\rbrac{x,y}$ acts as attractive potential.
In the photon BEC setup, the mirrors are placed close enough to each other to reach a low $q\approx 10$ at a cut-off wavelength at $\lambda_\text{cut}=\SI{580}{\nano\meter}$, resulting in a mechanical mirror distance of $D_0=\frac{q\lambda_\text{cut}}{2n}\approx\SI{2}{\micro\meter}$.

\subsection{Box potential}
\begin{figure}
    \centering
    \includegraphics[width=\linewidth]{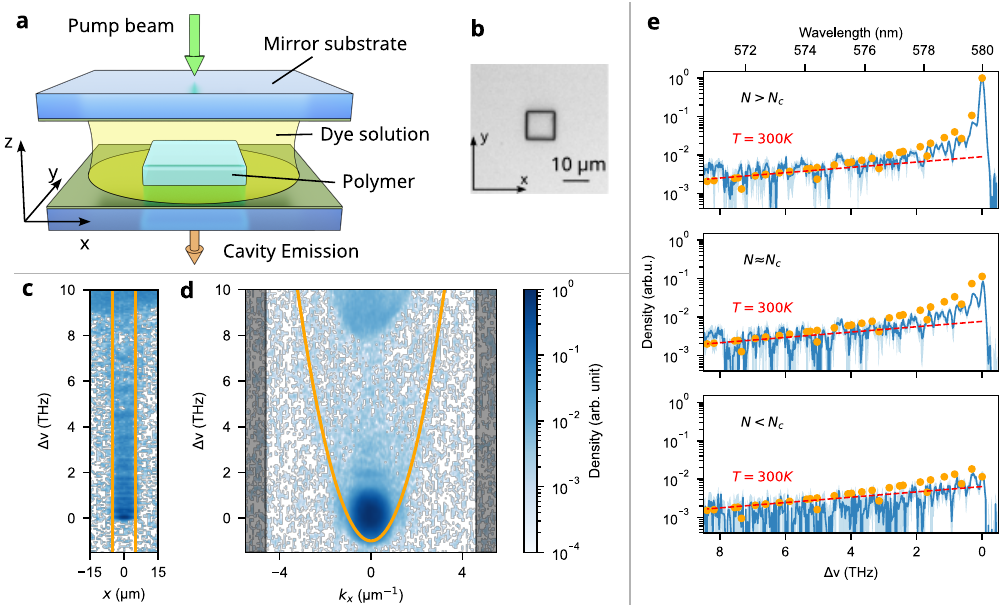} % sample 20240214 (box)
    \caption{
        Photon gas spectroscopy of a box potential. 
        \textbf{a} Sketch of the cavity in the Photon BEC setup. A polymer structure has been direct laser written on one of the cavity mirrors.
        \textbf{b} Microscope image of a polymer box potential.
        \textbf{c} Position space spectrum of the light from the cavity for a box potential.
        The light is confined to the width of the box marked by the straight orange lines.
        \textbf{d} Momentum space spectrum of the light from the cavity for a box potential.
        The orange parabola shows the dispersion relation of a free particle with a mass of $m_\text{ph}$.
        The grey shaded areas mark the limits of our measurement setup in momentum space due to the numerical aperture of the imaging objective.
        \textbf{e} Integrated position space spectrum (along $x$) of the cavity fluorescence for a box potential below ($N=55\pm2$), around ($N=144\pm3$) and above ($N=469\pm7$) the critical particle number$N_\text{c}$.
        The red dashed line shows an thermal distribution $\exp{(-h\Delta\nu/\kB T)}$ and the orange points show the expected relative photon density based on the mode degeneracy and the Bose-Einstein distribution for $T=\SI{300}{\kelvin}$.
        }
    \label{fig:box}
\end{figure}

As depicted in sketch Fig.~\ref{fig:box}~\textbf{a} the polymer structures are printed directly on one of the cavity mirrors and then examined in a %common 
standard photon BEC setup \cite{karkihalli_umesh_dimensional_2024}.
In order to demonstrate that we can indeed manipulate the potential with the polymer structures and that the photon gas can condense in these structures, we first fabricate a box potential with an edge length of \SI{10}{\micro\meter}.
A microscope image of such a polymer structure is depicted in Fig.~\ref{fig:box}~\textbf{b}.
The light from the cavity in position and momentum space is spectroscopically resolved as shown in Fig.~\ref{fig:box}~\textbf{c} and \textbf{d}, respectively.
The experimental setup to measure these spectra is explained in the method section.
In position space (Fig.~\ref{fig:box}~\textbf{c}), one observes that all the light is confined to the width of the potential.
In momentum space (Fig.~\ref{fig:box}~\textbf{d}), the light is confined by the dashed parabola given by the dispersion relation of a massive particle of mass $m_\text{ph}$.
In both cases, the mode spectrum follows the predicted shape of a 2D bosonic gas in a box potential\cite{Busley.2022}.
When the photon number is increased, the macroscopic occupation of the ground state, characteristic of the phase transition to a BEC, can also be observed, as Fig.~\ref{fig:box}~\textbf{e} shows. While for a homogeneous gas no true phase transition is expected in two dimensions, the finite size of the potential restores the characteristic macroscopic population of the ground mode \cite{Busley.2022}. 
The distribution well matches a Bose-Einstein distribution at $T=\SI{300}{\kelvin}$, confirming that the photon gas is in thermal equilibrium with the environment. 

Around $\SI{9}{\tera\hertz}$, an increase in intensity is observable in Fig.\ref{fig:box}~\textbf{b} and \textbf{c}, indicating the depth of the potential. Above this energy the modes are no longer bound to the box potential, and a larger mode density of free continuum modes can be excited.
This is above the thermal energy, which corresponds to a frequency of $\nu_\text{th}=\kB T/h=\SI{6.25}{\tera\hertz}$ for $T=\SI{300}{\kelvin}$.
Based on the potential depth of $h\times\SI{9}{\tera\hertz}$, we infer the height of this polymer structure $h_\text{S}$ to be around $\SI{475}{\nano\meter}$.

% Comparison with other structuring techniques
With this polymer structure, the optical path length of the cavity is locally increased by $h_\text{S}\nicefrac{\Delta n}{n}\approx\SI{36}{\nano\meter}$.
Laterally, this increase occurs over a distance that is determined by the radius of the voxel in the DLW-process which is $\approx\SI{50}{\nano\meter}$ and thereby far smaller than the wavelength of the photons in the cavity.
From the known voxel size, we expect the slope of our potential to be of order $\nicefrac{\Delta z}{\Delta x}\hat{=}\nicefrac{\SI{36}{\nano\meter}}{\SI{50}{\nano\meter}}=0.72$, i.e. of order one, which exceeds the slopes obtained by other techniques (thermal expansion layer reaches $\nicefrac{\SI{50}{\nano\meter}}{\SI{3.6}{\micro\meter}}=0.014$ \cite{Kurtscheid.2020}, milling of the mirror substrate $\nicefrac{\SI{230}{\nano\meter}}{\SI{2.5}{\micro\meter}}=0.092$ \cite{Trichet:15}). To verify that indeed the potential walls are steep, in the following we have investigated two tunnel-coupled micropotentials, where the lateral size determines the tunnel rate.

\subsection{High coupling rate in a double well potential}

\begin{figure}
    \centering
    \includegraphics[width=\linewidth]{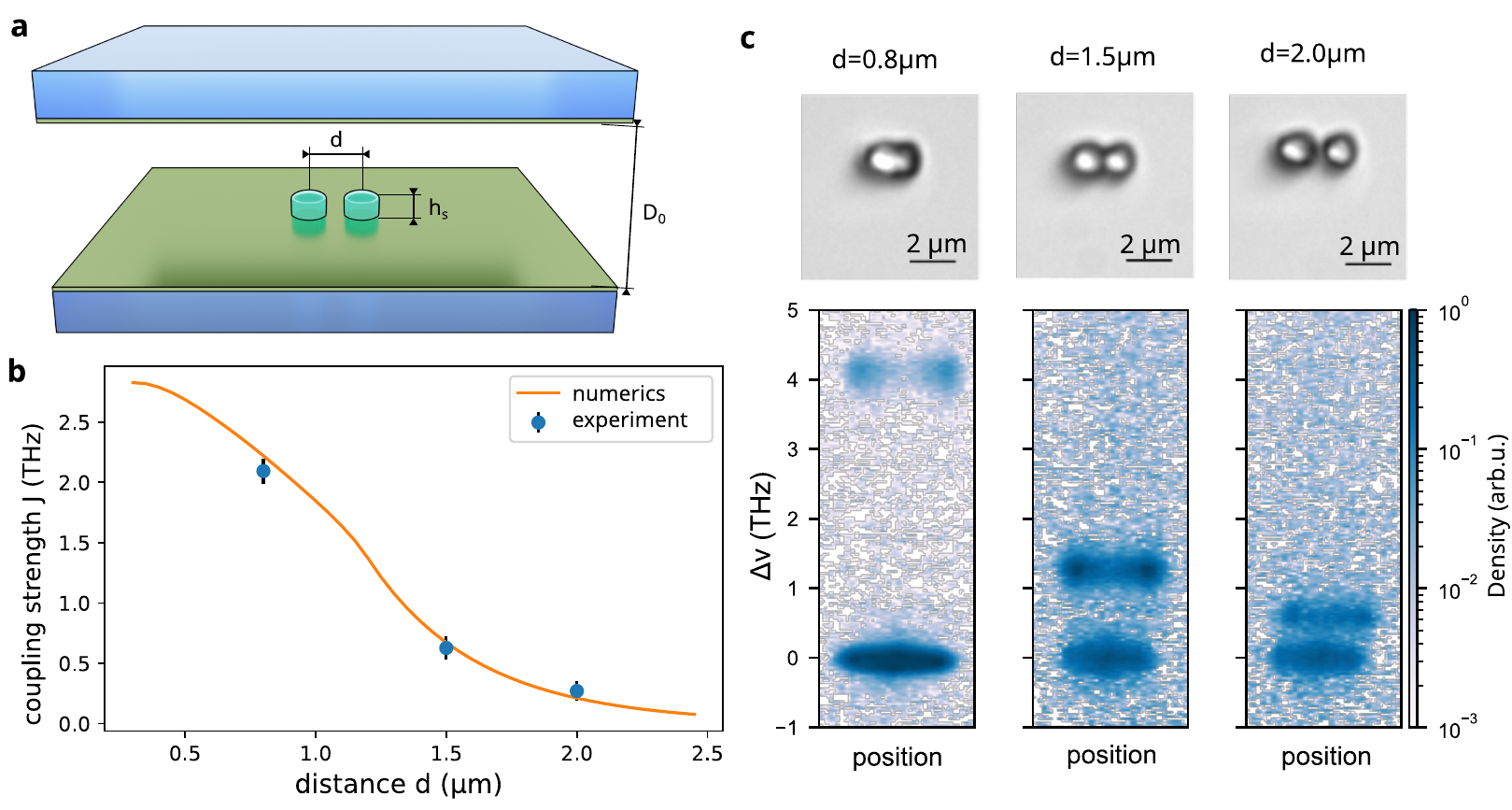} % sample 20240221 (double well)
    \caption{
        \textbf{a} Sketch of the cavity setup for a double well potential. 
        With a mechanical mirror distance $D_0$, a height of the polymer structure $h_\text{S}$ and a center-to-center distance between the pillars of $d$.
        \textbf{b} Dependence of the couplings strength over center-to-center distance $d$ of the two pillars. The solid orange line shows the numerically predicted strength, where $r=\SI{0.6}{\micro\metre}$, $h_s=\SI{0.6}{\micro\metre}$ and $q=10$. 
        \textbf{c} Position space spectrum of the cavity fluorescence for a cavity with two coupled pillars for different center-to-center distances $d$. 
        With increasing distance the mode splitting $\Delta\nu$ ($=2J$) decreases.
        On the top microscope images of the corresponding polymer structures are shown.
        }
    \label{fig:double_well}
\end{figure}

To study lattice physics with photons, one ideally requires structures where the tunnel coupling between the lattice sites is the dominant energy scale. To investigate this, we create double well potentials with varying tunnel coupling.
For this purpose, we print two circular pillars next to each other, as sketched in Fig.~\ref{fig:double_well}~\textbf{a}.
The radius of the pillars is chosen at $r=\SI{0.6}{\micro\meter}$ such that only one mode is bound in the potential of an individual pillar.
To vary the coupling strength $J(d)$, the distance from center to center between the pillars $d$ was adjusted.
This system of two coupled modes, one in the right pillar $\psiR$ and one in the left pillar $\psiL$ at the same energy $E_0$, can be described by a set of coupled mode equations \cite{Dung.2017}:
\begin{equation}
    \text{i}\hbar\partial_t
    \begin{pmatrix}
        \psiL\\
        \psiR
    \end{pmatrix}
    =
    \begin{pmatrix}
        E_0 & -J\\
        -J & E_0
    \end{pmatrix}
    \begin{pmatrix}
       \psiL\\
        \psiR 
    \end{pmatrix}.
\end{equation}
This system has two eigenmodes.
In the symmetric eigenmode $\psiS=\rbrac{\psiL+\psiR}/\sqrt2$ the light in both wells has the same phase.
Due to coupling, the energy of this mode is reduced to $E_\text{S}=E_0-J$.
The energetically next mode is the antisymmetric mode $\psiA=\rbrac{\psiL-\psiR}/\sqrt{2}$ with an energy of $E_\text{A}=E_0+J$, 
Due to the phase shift of \SI{180}{\degree} between the pillar modes, the light destructively interferes between the pillars, causing the characteristic double-lobe intensity pattern \cite{Kurtscheid.2019}.

Since we can resolve the cavity modes energetically in the experiment, we can directly measure the coupling strength in frequency units as half the frequency difference $\Delta\nu$ between the modes $\psiS$ and $\psiA$, as displayed in Fig.~\ref{fig:double_well}~\textbf{c}.
Here, we must note that our resolution in the position space is actually not high enough to resolve the double-lope shape of $\psiA$.
To make the two modes more distinguishable, we slightly defocussed from the position space focal plane, which enhances the double lope shape of $\psiA$.
The measurements agree well with numerical simulations, as shown in Fig.~\ref{fig:double_well}~\textbf{b}.
In the plotted simulation results of the coupling strength, one can notice a slight kink at \SI{1.2}{\micro\meter}, the distance where the pillars touch and one transitions from the regime of two tunnel-coupled potentials to the regime of a single tapered potential. 
At this contact point we obtain a coupling rate of about $J\approx\SI{1.5}{\tera\hertz}$, and for the tapered regime we observe experimentally a coupling of $J\approx\SI{2}{\tera\hertz}$ at a distance of $d=\SI{0.8}{\micro\meter}$. 
Compared to previous works\cite{Dung.2017, Kurtscheid.2019,Kurtscheid_2025}, which achieved a mode splitting of up to $\Delta\nu\approx\SI{50}{\giga\hertz}$, we observe mode splitting two orders of magnitude larger due to the higher spatial resolution, where the steeper walls of our pillars allow us to decrease the spacing between individual sites.

\subsection{1D lattice structures}

\begin{figure}
    \centering
    \includegraphics[width=\linewidth]{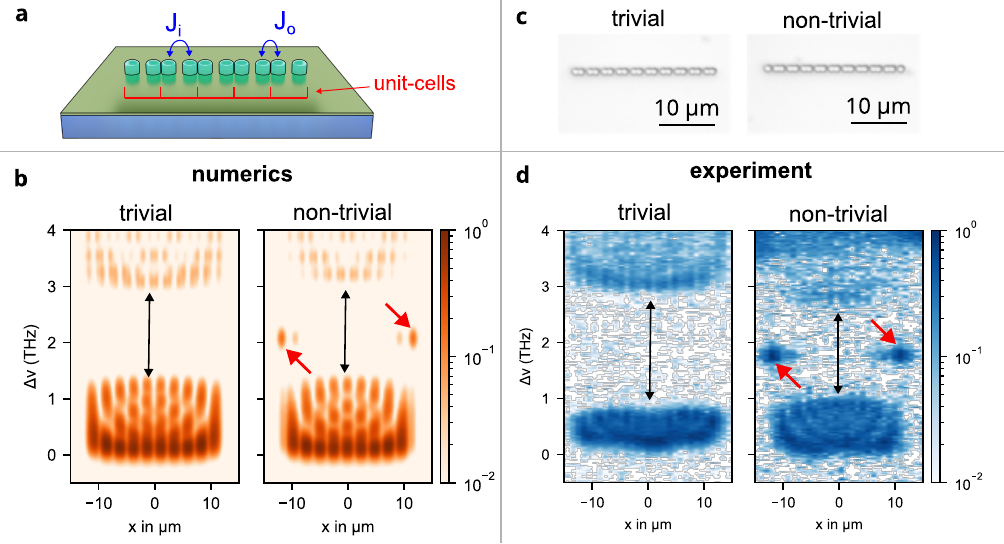} % sample  20241030 (SSH)
    \caption{
        \textbf{a} Sketch of the mirror surface structure imprinted on one of the cavity mirrors for a topological non-trivial SSH-array. 
        The coupling of sites inside $J_\text{i}$ and between $J_\text{o}$ the unit cells is controlled via their distance.
        \textbf{b} Predicted position space spectrum for the topological trivial (left) and topological non-trivial (right) structure. 
        In both cases exist a optical band gap (marked with black arrows) but only in the topological non-trivial case exist states in the center of the band gab which are localized at the ends of the array (marked with red arrows). 
        \textbf{c} Microscope images of the polymer structures.
        \textbf{d} Position space spectrum of the cavity fluorescence for an SSH array. 
        The experimentally observed spectra shows the predicted features, as the bandgap (marked with black arrows) and edge states (marked with red arrows).
        }
    \label{fig:1Darray}
\end{figure}

These results indicate that the photon lifetime in the cavity is sufficient such that a delocalized state over multiple sites can be formed. To demonstrate the ability to introduce major changes to the density of states, we choose to print a one dimensional Su-Schrieffer-Heeger (SSH)  chain \cite{Su_1979, Ozawa_2019} consisting of 20 sites (two sites per unit cell), coupled with $J_i$ within the unit cells and $J_o$ between the unit cells, as sketched in Fig.~\ref{fig:1Darray}~\textbf{a}.
Such an SSH chain has a band structure that spans $2\abs{J_i+J_o}$ consisting of two bands with a band gap of size $2\abs{J_i-J_o}$.
Two different topological phases can be identified depending on the coupling constants, the topological trivial for $J_i>J_o$ and the topological non-trivial for $J_o>J_i$.
Only in the topological non-trivial phase exist two states that have an energy in the center of the band gap that are localized at the edges of the array.

To predict the position space spectrum, the eigenmodes are numerically calculated, their intensity profile scaled in amplitude with a thermal distribution and shifted proportional to the corresponding wavelength.
This numeric prediction is shown in Fig.~\ref{fig:1Darray}~\textbf{b}.
In the topological trivial and the non-trivial case, an optical band gap of $\approx\SI{2}{\tera\hertz}$ (marked with black arrows) can be seen. Only in the topological non-trivial case two states in the center of the band gap are found, which are localized at the edges of the array (marked with red arrows), as expected. 

Microscope images of the polymer structures for the accompanying experiment are shown in Fig.~\ref{fig:1Darray}~\textbf{c} and the position space spectrum with these structures is shown in Fig.~\ref{fig:1Darray}~\textbf{d}.
The spectrum shows the presence of two bands, which are clearly separated, and also the substructure from the finite lattice size is clearly visible. For the case of the non-trivial lattice, the midgap edge states are clearly visible, localized at the two outermost pillars of the lattice. The position space spectrum agrees well with the numeric prediction and the characteristic features such as the bandgap and the existence of edge states are clearly identifiable.

This demonstrates that we can indeed create a strongly and coherently coupled photon gas over at least 20 sites.%, but this number of lattice sites is in our experience not even close to the limit. 
We want to emphasize that we did not study thermalization of photons into the lattice, for the presented results the pumping was adjusted to make the excited modes better visible, and thus we do not expect a thermal distribution within the modes. Correspondingly, the experiment can be understood as a proof-of-principle experiments to show the feasibility to prepare large coupled lattices for photons.

\subsection{Potentials with curvature}

\begin{figure}
    \centering
    \includegraphics[width=\linewidth]{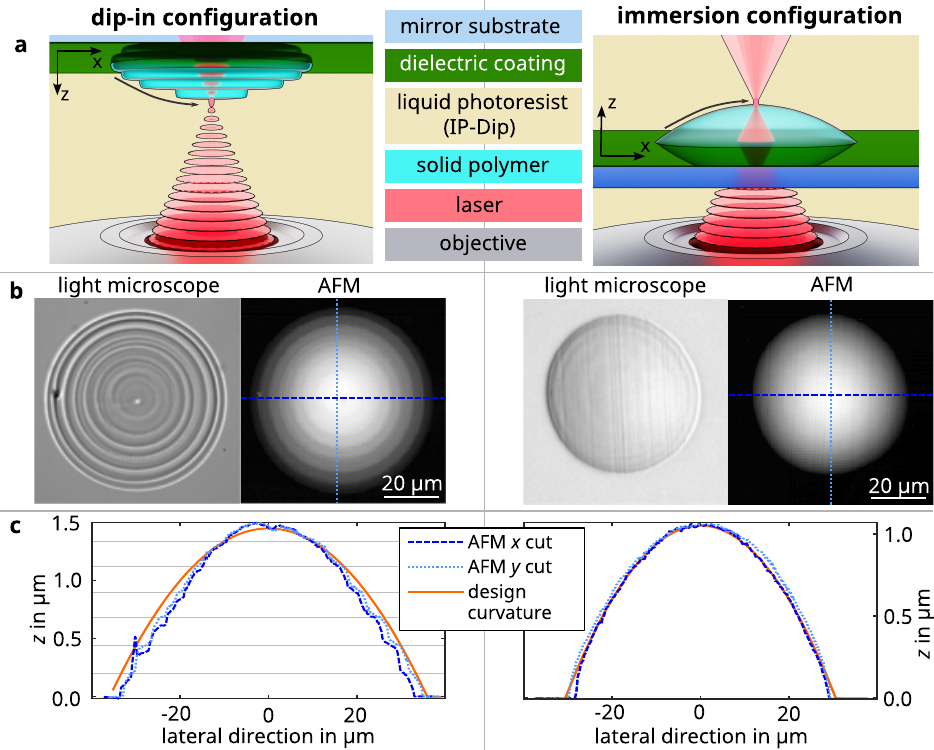}
    \caption{Dip-in configuration (left column) and immersion configuration (right column) in comparison.
    \textbf{a} Sketch of the used direct laser writing configurations. 
    In the dip-in configuration, reflection from the dielectric coating causes self-interference of the laser focus, which causes written structures to have discrete steps in height. 
    For the immersion configuration, a large part of the mirror substrate must first be removed. 
    On the other hand, there is no self-interference of the focus and structures with a smooth curved surface can be written.
    \textbf{b} Microscope image and Atomic force microscope measurement of the surface of an isotropic paraboloid written in the two configurations, respectively. 
    \textbf{c} Sections through the AFM data along the $x$ axis (blue dashed) and the $y$ axis (light blue dotted) compared to a parabola with the curvature programmed for the 3D print (orange straight). The parallel gray lines in the bottom left plot have a distance of \SI{230}{\nano\meter}.}
    \label{fig:DLW}
\end{figure}

Until now, we presented only potential structures that have a constant height.
Due to the fact that we are using 3D microprinting, we can also fabricate polymer structures with a position-dependent height $h_\text{s}(x,y)$, as we have demonstrated with paraboloids in Ref.~\cite{karkihalli_umesh_dimensional_2024}.
However, doing so requires a few more and slightly different fabrication steps to circumvent an inherent artifact of the structuring method.
For the flat structures, we have used the dip-in configuration for the DLW process. 
For curved structures, we have to switch to the immersion configuration.
Both DLW configurations are sketched in Fig.~\ref{fig:DLW}~\textbf{a}.

% sample 259 (paraboloid with steps) & 347 (paraboloid without steps)

% what is the problem (dip-in)
In the dip-in configuration, the photoresist serves directly as the immersion medium of the objective, and the structure is written "upside down" on the mirror. 
No special preparations are required for this configuration, but the reflection of the writing laser on the dielectric coating causes self-interference in the focus.
This self-interference in the focus causes discrete steps in the written structure, even if the trajectory of the writing laser had performed a continuous uniform movement in the $z$ direction.
An example of this effect is shown in the left column of Fig.~\ref{fig:DLW}~\textbf{b} and \textbf{c} in an AFM measurement of a paraboloid structure printed in a dip-in configuration.
The steps have a height of $\approx\SI{230}{\nano\metre}$ similar to the ones observed in Ref.~\cite{DLW_nanowire}.
Although this discretization of the height of structures is undesirable when curved surfaces are required, it can also be used as an advantage for flat structures, where the height of the potentials should be as constant as possible. 

% immersion configuration
However, the effect of self-interference can be completely avoided by using the immersion configuration. 
During the writing process, the laser is focused through an immersion medium and the mirror substrate in the photoresist (see Fig.~\ref{fig:DLW}~\textbf{a}). 
The beam is partially reflected on the dielectric coating back into the substrate, and therefore, reflection can \underline{not} interfere with the focus.
However, in this configuration, the working distance of the objective has to be greater than the substrate thickness and the maximal structure height, which in our case is \SI{360}{\micro\metre} (Plan-Apochromat 63$\times$/1.4 Oil DIC, Zeiss).
Therefore, most of the substrate from the backside of the cavity mirror has to be removed, which is done manually with sandpaper. 
Afterwards, the rough side is polished with fiber polishing / lapping film (LF30D, Thorlabs), until the major scratches are gone.
Minor scratches are less of a problem since they are filled up with a medium with a refractive index close to glass in the following steps anyway.
For the DLW process this medium is the immersion medium and for the following experiments with the photon BEC the medium is clear nail polish with which the mirror piece is glued on a 1~inch glass substrate for easier handling.
The structures produced in this way are very close to the programmed surfaces, which was verified by atomic force microscopy measurements (see right column of Fig.~\ref{fig:DLW}~\textbf{c}). We find a maximum deviation of about \SI{100}{\nano\metre} between design and measured surface\cite{karkihalli_umesh_dimensional_2024}.

\section{Methods}

% writing programm
For the DLW-process the commercial DLW-system \textit{Photonic Professional GT} (\textit{Nanoscribe}) is used.
The structures were written by moving the substrate only with piezoelectric actuators relative to the focal point of the objective. 
The piezoelectric actuators can move in all three dimensions, which allows us to correct for the tilt of the substrate.
The writing trajectory follows parallel lines with a line distance of \SI{100}{\nano\metre}. 
At each point, the $z$ position is adjusted according to the height of the desired structure and the tilt of the substrate.
For every structure, only the top surface layer is written.

% writing parameter
A laser intensity 1.5 times the polymerization threshold at a writing speed of \SI{100}{\micro\metre/\second} is used. 
For the dip-in configuration the polymerization threshold is at a laser power of 20\% while for the immersion configuration it is at 35\% due to the reflection from the dielectric coating. 
Here, a laser power of 100\% refers to a laser intensity of \SI{68}{\milli\watt} at the entrance pupil of the 63$\times$ focusing objective of the DLW system. 
For all polymer structures, the photoresist IP-Dip (\textit{Nanoscribe}) is used. 
In the immersion configuration, IP-Dip is also used as immersion medium between the substrate and the objective.
After the writing process, excess photoresist is removed by developing the sample in PGMEA (Propylenglycolmonomethyletheracetat, CAS:108-65-6) and isopropanol each for \SI{20}{\minute} and blow drying the sample with a nitrogen gun.
The air stream of the nitrogen gun was directed almost parallel to the mirror surface to reduce the number of dust particles sticking to the surface that have been accelerated perpendicular to the mirror surface by the air stream.

% structure recycling
To restructure a mirror, the mirror is first put in acetone.
This helps to loosen the polymer structures from the surface and, in the case of immersion configuration, it completely dissolves the nail polish, which detaches the mirror piece from the 1~inch~substrate. 
After that, the mirror surface can be cleaned with acetone, isopropanol, and lens cleaning wipes and then used again as the substrate for the DLW-process.

\begin{figure}
    \centering
    \includegraphics[width=\linewidth]{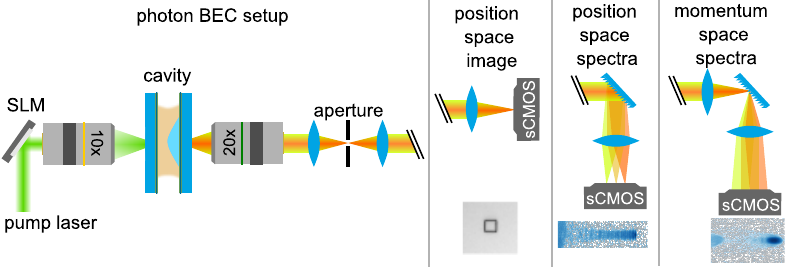}
    \caption{Schematic sketch of the photon BEC setup. 
    The pump beam is structured using a spatial light modulator (SLM) before entering the cavity via an objective. 
    The light emitted by the cavity is collected via a second objective on the other side of the cavity. 
    A lens is used to image the cavity in the focal plane of the lens, where an aperture can be used to block light modes that are not confined by the potential. 
    From there on, the light can be directed to various imaging setups to get the position space image, the position space spectra or the momentum space spectra. 
    }
    \label{fig:PBECSetup}
\end{figure}

% PBEC details
For the photon BEC setup, a \SI{1}{\milli\mol/\liter} dye solution of Rhodamine 6G dissolved in ethylene glycol is placed within the cavity consisting of the structured and another unstructured plane mirror.
At the wavelength of the photon gas around \SI{580}{\nano\meter}, IP-Dip has a refractive index of 1.55 \cite{Gissibl:17} and ethylene glycol of 1.44 creating a refractive index difference of $\Delta n=0.11$.
The \SI{532}{\nano\meter} pump light is structured with a spatial light modulator (SLM) and coupled into the cavity via an objective and through the unstructured mirror. The photon gas is observed through the structured mirror and a second objective as sketched in Fig.~\ref{fig:PBECSetup}.
A variable rectangular aperture in one of the position space image planes is used to block most of the light that is not coming from the potential.
This reduces the overlap in the position space spectrums of free-space modes with the modes bound by the potential, which we want to investigate.
From the aperture on, the light can be imaged onto a camera to obtain a position space image. 
Alternatively it can be spectrally resolved by reflecting it off a grating and then be imaged onto a camera to obtain the position space spectra, where for our slitless spectrometer the spectral information is convoluted with the corresponding mode shape \cite{karkihalli_umesh_dimensional_2024}. 
With an additional lens in the beam path, the momentum space spectra of the light can be imaged by reflecting it off a grating and then imaging the back focal plane onto a camera.

% simulations
To simulate the expected energies and the electric field distributions of the cavity eigenmodes, we used the python package PyPBEC~\cite{PyPBEC_package_git} that was developed and openly published by the authors of Ref.~\cite{PyPBEC_package_paper}.

\section{Conclusions}

We have presented a new structuring technique to create sharp (steep) and deep potentials for photon BECs.
With this technique, we have prepared various potential landscapes for the photon gas. In a box potential, we have observed a thermal occupation of modes, with a macroscopic population of the ground mode for increasing photon numbers. Using double well potentials, we have demonstrated  large coupling rates that are an order of magnitude higher than those in previous works, showing that we can confine the photon gas to sharp and arbitrary-shaped potentials.
As a proof-of-principle to show the feasibility of lattice physics with photons, we have investigated the SSH model in a lattice with 20 sites. %we have demonstrated for the first time that the photon gas of multiple tens of
%sites can be coupled together to investigate lattice phenomena.
Due to these features and the high resolution of this new structuring approach, we expect that this technique will be of great interest and widely adopted in the community. Prospects include the study of vortices in driven-dissipative photon condensates \cite{GladilinWouters_2020, Krauss_2025}, the investigation of localization in coupled chains with thermooptic interactions \cite{Strinati_2022} or the observation of transport phenomena such as the bosonic skin effect in tilted lattices \cite{Garbe_2024}.

\section{Acknowledgements} \par %delete if not applicable))
We acknowledge valuable discussions with E. Stein, J. Krauß and A. Pelster, as well as M. Weitz, J. Schmitt, A. Redmann and L. Kleebank. This work has been supported  by the Deutsche Forschungsgemeinschaft through CRC/Transregio 185 OSCAR (project No.\ 277625399).

\bibliography{DLW_phBEC}{} % Produces the bibliography via BibTeX.
\bibliographystyle{unsrt}

\end{document}